\documentclass[graybox]{svmult}

\usepackage{physics}

\usepackage{type1cm}        
\usepackage{makeidx}         
\usepackage{graphicx}        
\usepackage{multicol}        
\usepackage[bottom]{footmisc}

\usepackage{newtxtext}       %
\usepackage{newtxmath}       

\usepackage{soul}

\usepackage{color}
\usepackage{mhchem}
\makeindex             
                       
\newcommand{\ii}{\mathrm{i}}
\newcommand{\ee}{\mathrm{e}}
\newcommand{\dx}{\dd x}
\newcommand{\dy}{\dd y}

\begin{document}

\title*{Purcell effect in PT-symmetric waveguides}
\author{Alina Karabchevsky, Andrey Novitsky and Fyodor Morozko}
\institute{Alina Karabchevsky \at School of Electrical and Computer Engineering, Ben-Gurion University of the Negev, Beer-Sheva, 8410501, Israel, \email{alinak@bgu.ac.il}
\and Andrey Novitsky \at Department of Theoretical Physics and Astrophysics,
Belarusian State University, Nezavisimosti Avenue 4, 220030 Minsk, Belarus \email{andreyvnovitsky@gmail.com} 
\and Fyodor Morozko \at School of Electrical and Computer Engineering, Ben-Gurion University of the Negev, Beer-Sheva, 8410501, Israel, and \at Department of Theoretical Physics and Astrophysics,
Belarusian State University, Nezavisimosti Avenue 4, 220030 Minsk, Belarus \email{fyodormorozko95@gmail.com}}

%
%
\maketitle

\abstract{This chapter overviews the principles of the spontaneous emission rate increase, that is the Purcell effect, in relation to the photonic parity-time (PT) symmetry. Being focused on the system of coupled PT-symmetric optical waveguides, we consider behaviors of the Purcell factor in PT-symmetric and broken-PT-symmetric regimes. Surprisingly, exceptional points in a coupled waveguide do not influence on the Purcell factor. 
}

\section{Introduction}

By exploring the interplay between loss and gain as well as the coupling mechanism in waveguide-emitter systems, one can generate and control light on a chip. This chapter introduces the underlying physics of Purcell effect for emitters in PT-symmetric waveguides. In general, physical world exhibits symmetries lying behind the conservation laws of physics. They help to control the structure of matter and define interactions. The laws of physics are required to be invariant under changes of redundant degrees of freedom dictated by the symmetries. There are several fundamental symmetries including the charge conjunction or \textit{C symmetry} for a particle and its anti-particle, parity or \textit{P symmetry} for a system and its mirror image and time reversal or \textit{T symmetry} for the time running forward and backward. Despite the fact that the laws of physics are dictated by symmetries, it is the symmetry breaking that creates nontrivial physics by lifting the degeneracies. A number of intriguing properties in photonics are related to the PT-symmetry usually described by non-Hermitian systems. Non-Hermitian Hamiltonians possessing parity-time (PT) symmetry that is the symmetry with respect to the simultaneous coordinate and time reversal \cite{ref:zyablovsky2014pt}. There is still a debate whether PT-symmetry is a fundamental feature or shares common properties with naturally occurring symmetries. Also questionable the phenomenon of phase transition and it is important to understand the spectral degeneracies induced by PT-symmetry named exceptional points (EP) which is a point in parameter space at which phase transition occurs.

\begin{figure}[t]
\centering
\includegraphics[scale=1.1]{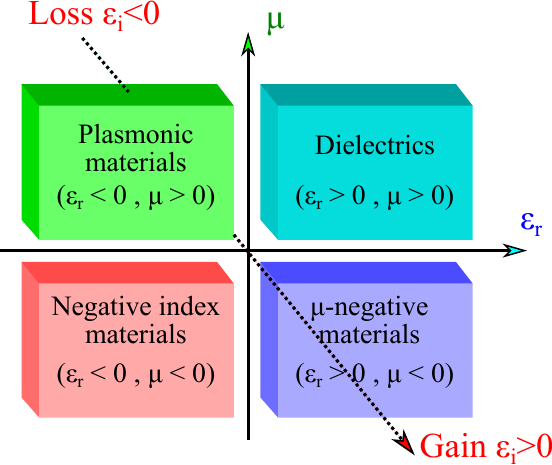}
%
%
\caption{ Schematics of materials characterisation in terms of loss and gain.
}
\label{fig:PTsym}       
\end{figure}

 Controlling the magnetic permeability $\mu$ and the real part of the dielectric permittivity $\varepsilon_r$ has enabled novel functionalities. PT symmetry and non-Hermitian photonics open new possibilities by controlling the imaginary part of the dielectric permittivity ($\varepsilon_i$), and by considering gain and loss. Figure~\ref{fig:PTsym} schematically shows this interplay when characterising materials in terms of the gain and loss.

The Chapter is organized in the following way:
In Section~\ref{sec:pt_principles}, we introduce the principles of PT-symmetry.
PT-symmetric photonic devices are presented in Section~\ref{sec:devices} such as those based on coupled-mode theory, two-dimensional photonic waveguide lattices, multilayer structures, and microresonators.
Purcell effect in PT-symmetric waveguides is described in Section~\ref{sec:purcel_pt_wg}.
Eventually, the Section~\ref{sec:summary} summarises the Chapter and gives an outlook to future research.

\section{Principles of PT symmetry}
\label{sec:pt_principles}

In 1998 Bender and B\"otcher~\cite{ref:bender1998} have shown that quantum systems with non-Hermitian Hamiltonians can have entirely real spectra.
Such Hamiltonians are referred to as pseudo-Hermitian.
The first known class of pseudo-Hermitian Hamiltonians are PT-symmetric ones i.e. those commuting with $\hat{P}\hat{T}$ operator where $\hat{P}$ and $\hat{T}$ are correspondingly coordinate and time reversal operators
\begin{equation}
  \hat{P}\hat{T}\hat{H}=\hat{H}\hat{P}\hat{T}.
\end{equation}

Operator $\hat{P}$ changes sign of coordinates whereas $\hat{T}$ operator changes sign of time and performs complex conjugation~\cite{ref:zyablovsky2014pt}.
This means that PT-invariance of the Hamiltonian can be stated in the form
\begin{equation}
  \hat{H}(\hat{\vb{p}},\hat{\vb{r}},t)=\hat{H}^*(\hat{\vb{p}},-\hat{\vb{r}},-t).
\end{equation}

For Hamiltonians of the form
\begin{equation}
  \hat{H}=\frac{\hat{\vb{p}}^2}{2m}+V(\vb{r}),
  \label{eq:particle_H}
\end{equation}
where $\hat{\vb{p}}$ is the momentum operator, $m$ is mass, and $V$ is complex potential, action of $\hat{P}\hat{T}$ operator results in the Hamiltonian
\begin{equation}
  \hat{P}\hat{T}\hat{H}=\hat{H}\hat{P}\hat{T}=\frac{\hat{\vb{p}}^2}{2m}+V^*(-\vb{r}).
  \label{eq:particle_PTH}
\end{equation}
Therefore, for the Hamiltonian~\eqref{eq:particle_H} to be PT-invariant it is needed that potential energy $V(\vb{r})$ satisfies the condition
\begin{equation}
  V(\vb{r})=V^*(-\vb{r}).
  \label{eq:pt_V}
\end{equation}
In other words, real part of the potential energy must be even function of coordinates whereas imaginary part must be odd function.

It can be shown that if the eigenfunctions $\ket{\psi_n}$ of the PT-symmetric Hamiltonian $\hat H$,
\begin{equation}
  \hat H \ket{\psi_n}=E_n\ket{\psi_n},
\end{equation}
with the corresponding eigenvalues $E_n$ are also eigenfunctions of the $\hat{P}\hat{T}$ operator
\begin{equation}
  \hat{P}\hat{T}\ket{\psi_n}\equiv\sigma_n\ket{\psi_n}
  \label{eq:PTpsi}
\end{equation}
with some (complex) eigenvalues $\sigma_n$,
the eigenvalues $E_n$ of the Hamiltonian are real.
Condition~\eqref{eq:PTpsi} is necessary and sufficient for eigenvalues of the Hamiltonian~\cite{ref:zyablovsky2014pt} to be real.
Hence, if the eigenvalues $E_n$ are real, the eigenfunctions are PT-symmetric and the system is
considered to be in PT-symmetric regime (phase).
Contrarily, if the eigenvalues are complex, the eigenfunctions are essentially not PT-symmetric and the system is in PT-symmetry-broken regime.

In contrast to Hermitian case,
eigenfunctions of pseudo-Hermitian Hamiltonians are not orthogonal under conventional inner product.
Instead, they obey more general biorthogonality relations.
Orthogonality can be re-established by modifying the inner product.
Discussion of quantum mechanics based on biorthogonal states is given in
\cite{ref:weigert2003,ref:mostafazadeh2010,ref:moiseyev2011a,ref:brody2016}.

\subsection{Phase transition in PT-symmetric systems}
\label{subsec:phase_transition}
One of the most intriguing features of PT-symmetric systems is a phase transition from the PT-symmetric to PT-symmetry-broken phase.
If the Hamiltonian of the system $\hat{H}(\hat{\vb{p}},\hat{\vb{r}},t,p)$ depends on some parameter $p$, the Hamiltonian can have real as well as complex eigenvalues being, either in PT-symmetric or non-PT-symmetric states, respectively.
When due to variation of the parameter $p$ the system's spectrum changes from real to complex and vice-versa,
one can study the phase transition related to the spontaneous breaking of PT symmetry. The point in parameter space $p=p_c$ at which phase transition occurs is named as an exceptional point (EP).
At the EP, both eigenvalues and eigenfunctions coalesce.

\subsection{PT-symmetry in optics}
\label{subsec:pt_in_optics}
Quantum-mechanical concept of the PT symmetry can be realised in optics due to the fact that Maxwell's equations in case of two- and one-dimensional photonic structures can be reformulated into an equation formally coinciding with the Schr\"odinger equation.

With light propagation along these structures it is convenient to introduce the so called slowly varying envelope (SVE) field, where most of the electromagnetic field variation is extracted by defining a suitable selected reference propagation constant~\cite{ref:gines2015}.
Specifically, if the structure is invariant in $z$, SVE $\vb{e}$ for electric field is defined as
\begin{equation}
  \vb{E}(x,y,z)=\vb{e}(x,y,z)\ee^{-\ii k_0n_0 z}
  \label{eq:Eenvelope}
\end{equation}
where $k_0=\omega/c$ is the vacuum wavenumber and $n_0$ is the reference (background) refractive index.
SVE $\vb{h}$ for magnetic field is defined analogously as
\begin{equation}
  \vb{H}(x,y,z)=\vb{h}(x,y,z)\ee^{-\ii k_0n_0 z}.
  \label{eq:Henvelope}
\end{equation}
Within the slowly varying envelope approximation (SVEA) it is assumed that
\begin{equation}
  \pdv[2]{z}\mqty(\vb{e}\\\vb{h}) \ll 2k_0n_0\pdv{z}\mqty(\vb{e}\\\vb{h})
\end{equation}
and the second-order $z$-derivative terms are neglected.
Due to $z$-invariance of the structure transverse and longitudinal components of $\vb{e}$ and $\vb{h}$ decouple.
The transverse components of SVE fields $\vb{e}_t$ and $\vb{h}_t$, hence, satisfy first-order equations with respect to the $z$ derivative.
The above can be summarized in a Schr\"odinger-like equation
\begin{equation}
  \ii\pdv{z}\ket{\psi}=\hat H\ket{\psi},
  \label{eq:schroedinger_opt}
\end{equation}
for an optical state-vector $\ket{\psi}$ defined as
\begin{equation}
  \ket{\psi}=\mqty(\vb{e}_t\\\vb{h}_t).
  \label{eq:state_vector}
\end{equation}
$\hat H$ is an optical Hamiltonian governing the $z$-evolution of SVE fields.
Generally, $\hat H$ is represented by a $4\times4$ matrix joining
operators describing evolution of $\vb{e}_t$ and $\vb{h}_t$.
Explicit form of these operators found in~\cite{ref:gines2015}.
For waveguide structures with very small index contrast in both transverse directions equation \eqref{eq:schroedinger_opt} can be reduced to a scalar equation.
Within the scalar approximation the Hamiltonian $\hat H$ takes the form
\begin{equation}
  \hat H=\frac{1}{2k_0n_0}\left(\pdv[2]{x}+\pdv[2]{y} + V(x,y)\right).
  \label{eq:Hscalar}
\end{equation}
Quantity 
\begin{equation}
  V(x,y)=k_0^2\left(\varepsilon(x,y)-n_0^2\right)
  \label{eq:opticalV}
\end{equation}
can be associated with a potential of the Sch\"odinger equation.
From the condition of the PT symmetry in quantum mechanics $V(x,y) = V^*(-x,-y)$ we arrive at the similar condition in optics $\varepsilon(x,y)=\varepsilon^*(-x,-y)$.
Therefore, in optical systems the PT symmetry can be established by judiciously incorporating gain and loss.
Thus, the refractive index profile now plays the role of the complex potential.

\subsection{Inner product for PT-symmetric optical systems}
We define the inner product as a cross product of the bra-electric and ket-magnetic fields integrated over the cross-section $z={\rm const}$:
\begin{equation}
  \braket{\phi_1}{\phi_2} \equiv \intop_{}
  \left( \vb{E}_{1}\times\vb{H}_{2} \right) \cdot \vu{z} \dx\dy
  \label{eq:inner_product}
\end{equation}
Such a definition is justified by the non-Hermitian nature of PT-symmetric systems.
In the above and following relations we can drop $t$ subscripts because $z$
component of the vector products depends only on transverse components and $\left( \vb{E}_{t,1}\times\vb{H}_{t,2} \right) \cdot \vu{z}=\left( \vb{E}_{1}\times\vb{H}_{2} \right) \cdot \vu{z}$.

It is well known that the modes of Hermitian systems are orthogonal in the sense
\begin{equation}
  \braket{i}{j^*}=\intop \left( \vb{e}_{i}\times\vb{h}_{j}^{*} \right) \cdot \vu{z}\dx\dy \sim \delta_{ij},
  \label{eq:orth_conventional}
\end{equation}
where $\delta_{ij}$ is the Kronecker delta.
Here and below $\ket{\psi^*}=(\vb{e}^*_t,\vb{h}^*_t)^T$.
Relationship~\eqref{eq:orth_conventional} is often referred to as power orthogonality, because $\frac12\Re\braket{i}{i^*}$ is the power carried by the mode $\ket{i}$.
However, the loss and gain in the non-Hermitian systems break power orthogonality.
In this case, one should use a non-conjugate inner product~\cite{ref:snyder1984,ref:svendsen2013,ref:wu2019} bringing us to the orthogonality relationship
\begin{equation}
  \braket{i}{j} = \intop \left( \vb{e}_{i}\times\vb{h}_{j} \right) \cdot \vu{z} \dd x \dd y = 2 N_i \delta_{ij},
  \label{eq:orth}
\end{equation}
where $N_i$ is a normalization parameter.
We want to stress that orthogonality relation~\eqref{eq:orth} is valid not only for PT-symmetric but for arbitrary non-Hermitian systems.

Forward and backward transverse modal fields $\vb{e}_{t,i}$ and $\vb{h}_{t,i}$ satisfy the symmetry relations
\begin{equation}
  \vb{e}_{t,-i}=\vb{e}_{t,i},\qquad\vb{h}_{t,-i}=-\vb{h}_{t,i}
  \label{eq:forward_backward}
\end{equation}
both in the case of Hermitian and non-Hermitian systems.

This means that the inner product of the modes also meets the symmetry relations for its bra- and ket-parts:
\begin{align}
  \braket{-i}{j}&=\braket{i}{j},\\
  \braket{i}{-j}&=-\braket{i}{j}.
  \label{eq:forward_backward_braket}
\end{align}

\subsection{Petermann factor}
\label{subsec:petermann}
It is common to express non-orthogonality of the modes quantitatively in terms of Petermann factor~\cite{ref:petermann1979,ref:siegman1989,ref:berry2003,ref:yoo2011,ref:pick2017}.
Petermann factor is defined as the squared ratio between Hermitian and non-Hermitian norms.
In our notation Petermann factor $K_i$ of the mode $\ket{i}$ reads as
\begin{equation}
  K_i=\frac{\abs{\bra{i}\ket{i^*}}^2}{\abs{\braket{i}}^2}.
  \label{eq:petermann_factor}
\end{equation}
Petermann factor obviously equals to unity in Hermitian case since in this case transverse modal fields always can be rescaled to be real.
Hence, non-Hermitian norm is equal to the Hermitian norm and to the power carried by the mode.

\subsection{Eigenmodes of PT-symmetric optical systems}
To get some insight on the eigenstates of photonic PT-symmetric systems, let us analyze the system of two coupled waveguides using the coupled mode theory.
Coupled waveguides are the simplest systems proposed at the beginning of the era of optical PT symmetry \cite{ozdemir2019parity}.
As schematically shown in Fig. \ref{fig:Sect3}(a), they consist of gain and lossy waveguides having identical geometrical parameters at a distance $g$ one from another.
The waveguide can be either slab, rectangular, circular or gradient one, yet the physics behind the coupling mechanism is the same.

We express the total field in the coupled system in terms of the modes
$\ket{g}=(\vb{e}_{g,t},\vb{h}_{g,t})^T$ and $\ket{l}=(\vb{e}_{l,t},\vb{h}_{l,t})^T$ of isolated gain and loss waveguides
with corresponding $z$-dependent amplitudes $g$ and $l$ as
\begin{equation}
  \ket{\psi}=g(z)\ket{g}+l(z)\ket{l}.
  \label{eq:gl_ansatz}
\end{equation}
We assume that the overlap between the modes of isolated waveguides is negligible
(weak coupling condition), therefore, the modes are orthogonal and normalized as follows
\begin{align}
  \braket{g}{l}&=\braket{g}{l^*}=0,\\
  \braket{g}&=\braket{l}=1.
  \label{eq:gl_norm}
\end{align}

$\hat{P}\hat{T}$ operator converts the mode of isolated lossy waveguide to the mode of the isolated gain waveguide and vice versa, and so
\begin{subequations}
\begin{align}
  \hat{P}\hat{T}\ket{g} &= \ket{l},\\
  \hat{P}\hat{T}\ket{l} &= \ket{g}.
\end{align}
\label{eq:PTgl}
\end{subequations}

Coupled mode theory for optical PT-symmetric systems can be formulated on the basis of Lagrangian formalism \cite{ref:elganainy2007} or by using Lorentz reciprocity theorem~\cite{ref:chuang1987}.

Spatial evolution of amplitudes is governed by the system of coupled equations
\begin{equation}
  \ii\dv{z}\mqty(g\\l) =
  \mqty(\Re(\beta+\delta)-\ii\alpha/2&&\kappa\\
    \kappa&&\Re(\beta+\delta)+\ii\alpha/2)\mqty(g\\l)
  \label{eq:cmt_evolution}
\end{equation}
where $\beta$ is a propagation constant, $\kappa$ is a coupling coefficient,
$\delta$ is a correction to the propagation constant, $\alpha$ is an effective gain (or loss).
It can be shown that due to the weak coupling and relations \eqref{eq:PTgl}
the coupling constant $\kappa$ is real \cite{ref:elganainy2007,ref:chuang1987}.

Matrix in the right hand side of~\eqref{eq:cmt_evolution} is the matrix of the system's Hamiltonian in the basis $\ket{g},\ket{l}$.

\begin{figure}[htb]
  \centering
  \includegraphics[width=1.0\linewidth]{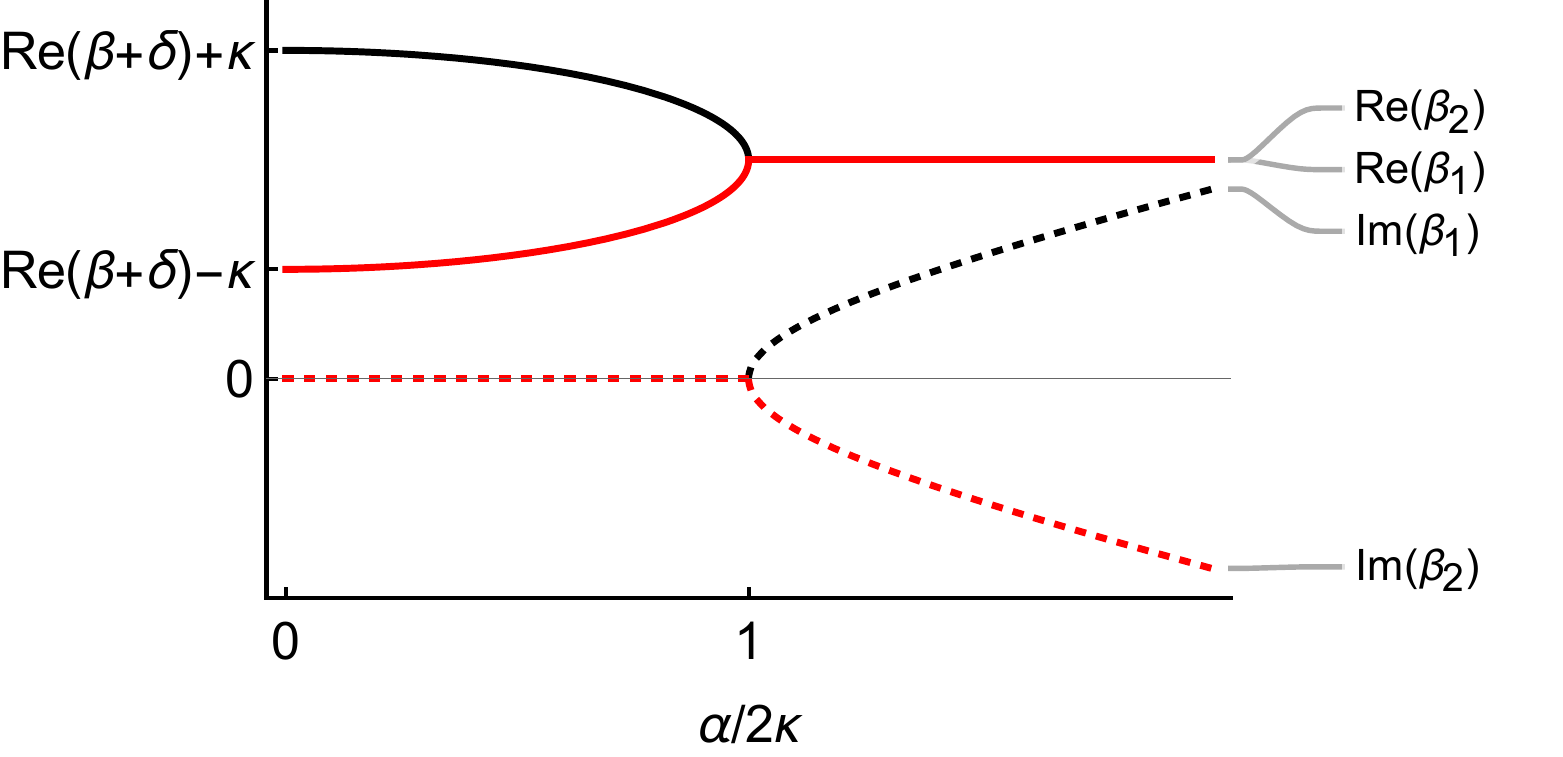}
  \caption{
  Eigenvalues of the coupled waveguide system's Hamiltonian versus
  non-Hermiticity parameter $\alpha/2\kappa$.
  Black curves correspond to the first supermode,
  red curves correspond to the second supermode.
  Solid curves correspond to real parts, dotted curves correspond to imaginary parts.}
  \label{fig:phase_diag}
\end{figure}

The eigenvalues of this Hamiltonian are the propagation constants of the sytem's eigenmodes.
They read as
\begin{equation}
  \beta_{1,2}=\Re(\beta+\delta)\pm\sqrt{\kappa^2-\alpha^2/4}.
  \label{eq:cmt_eigvals}
\end{equation}
Clearly, the system behaves differently depending on whether $\alpha/2$ is less or greater than $\kappa$.
When $\alpha/2$ is less than $\kappa$ both propagation constants are real.
When $\alpha/2$ is greater than $\kappa$ the eigenvalues constitute complex-conjugate pair
and one mode experiences gain whereas the other one experiences loss.
When $\alpha=\alpha_c=2\kappa$ modes degenerate.
Therefore the point $\alpha=\alpha_c$ corresponds to exceptional point (EP).
The situation when $\alpha$ passes through $\alpha_c$ is called the phase transition.
Phase diagram of a PT-symmetric coupled waveguide system in Fig.~\ref{fig:phase_diag} shows distribution of real and imaginary parts of system's eigenvalues.
It illustrates a typical picture of the phase transition in a PT-symmetric system.

In PT-symmetric regime, the eigenvalues can be written as
\begin{equation}
  \beta_{1,2}=\Re(\beta+\delta)\pm\kappa\cos\theta,
  \label{eq:eigvals_sym}
\end{equation}
where $\sin\theta=\alpha/2\kappa$.
With this parametrization supermodes take the form
\begin{equation}
  \ket{1,2}=\ket{g}\pm\ee^{\pm\ii\theta}\ket{l}.
  \label{eq:eigmodes_sym}
\end{equation}

It can be seen from \eqref{eq:PTgl} that the states \eqref{eq:eigmodes_sym} are indeed the eigenstates of the $\hat{P}\hat{T}$ operator
\begin{align}
  \hat{P}\hat{T}\ket{1}&=\ket{l}+\ee^{-\ii\theta}\ket{g}=\ee^{-\ii\theta}\ket{1},\\
  \hat{P}\hat{T}\ket{2}&=\ket{l}-\ee^{+\ii\theta}\ket{g}=\ee^{+\ii\theta}\ket{2}.
\end{align}

In the PT-broken regime, eigenvalues can be written as
\begin{equation}
  \beta_{1,2}=\Re(\beta+\delta)\pm\ii\kappa\sinh\theta,
  \label{eq:eigvals_broken}
\end{equation}
where $\cosh\theta=\alpha/2\kappa$.
Supermodes then read as
\begin{equation}
  \ket{1,2}=\ket{g}+\ii\ee^{\mp\theta}\ket{l}.
  \label{eq:eigmodes_broken}
\end{equation}

The eigenmodes in PT-broken regime are not longer the eigenstates of the $\hat{P}\hat{T}$ operator.
Instead, in this regime $\hat{P}\hat{T}$ operator relates $\ket{1}$ and $\ket{2}$ as follows
\begin{align}
  \hat{P}\hat{T}\ket{1}&=\ket{l}-\ii\ee^{-\theta}\ket{g}=\ii\ee^{-\theta}\ket{2},\\
  \hat{P}\hat{T}\ket{2}&=\ket{g}-\ii\ee^{\theta}\ket{g}=-\ii\ee^{\theta}\ket{1}.
\end{align}

\begin{figure}[t!b!]
  \centering
  \includegraphics[width=0.7\linewidth]{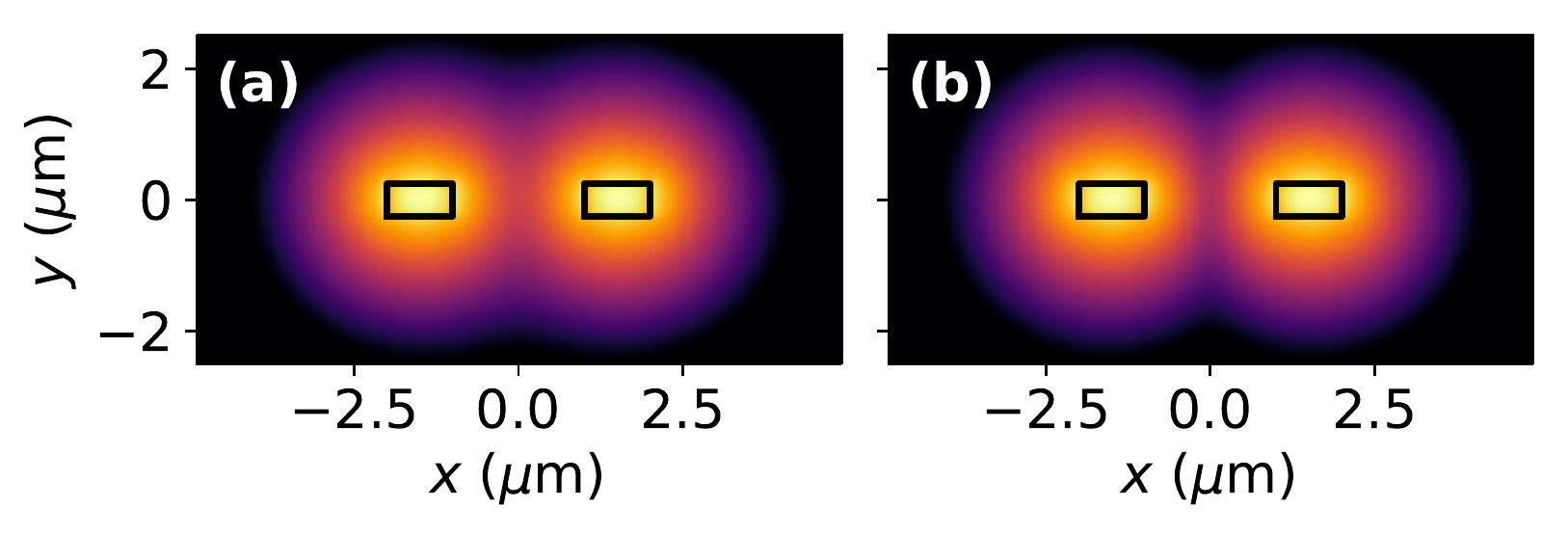}
  \caption{
  Mode profiles in PT-symmetric regime: (a) mode $\ket{1}$ and (b) mode $\ket{2}$.
  }
  \label{fig:modes_below_ep}
\end{figure}

\begin{figure}[t!b!]
  \centering
  \includegraphics[width=0.7\linewidth]{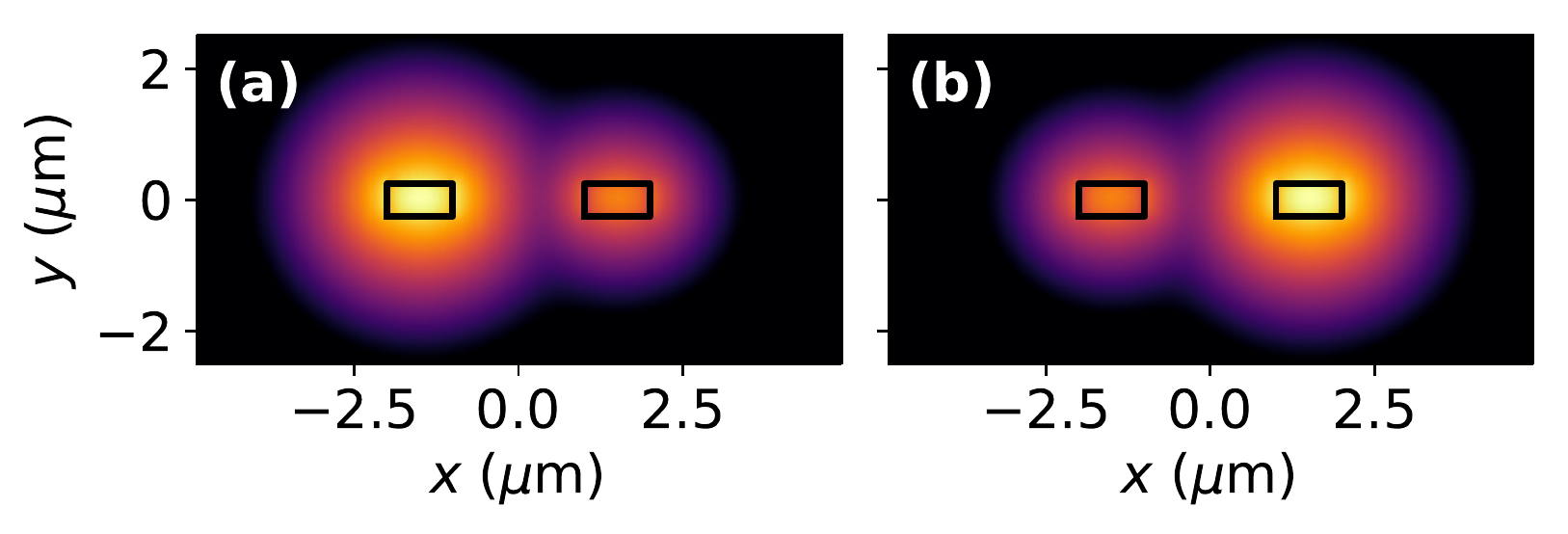}
  \caption{
  Mode profiles in PT-symmetry-broken regime: (a) mode $\ket{1}$ and (b) mode $\ket{2}$.
  }
  \label{fig:modes_above_ep}
\end{figure}

Typical mode profiles for the coupled waveguide system (see Fig.~\ref{fig:Sect3}(a)) in PT-symmetric and PT-broken regimes are shown in Fig.~\ref{fig:modes_below_ep}~and in Fig.~\ref{fig:modes_above_ep}.

When the parameter $\alpha$ approaches the value $\alpha_c=2\kappa$ corresponding to the EP,
eigenmodes $\ket{1,2}$ merge to become $\ket{g}+\ii\ket{l}$.
Interestingly, that the modes become self-orthogonal as $\braket{1}=\braket{2}=0$ at the EP.
Self-orthogonality is responsible for singularity of Petermann factor due to zero in denominator in Eqn.~\eqref{eq:petermann_factor}.

\section{PT-symmetric photonic devices}
\label{sec:devices}

Photonics is an excellent platform for experimental verification of the fundamental concept of the parity-time symmetry discussed earlier.
Novel photonic devices can be fabricated using several basic types of PT-symmetric structures, such as waveguides, multilayer systems and photonic crystals.
A number of remarkable applications of the PT symmetry have been proposed and well studied including unidirectional invisibility, lasing, sensing and coherent perfect absorption.
In this section, we overview the recent PT-symmetric photonic devices with the application perspective.

\subsection{Coupled waveguide systems}
\label{subsec:2}

\begin{figure}[t]
\centering
\includegraphics[scale=0.65]{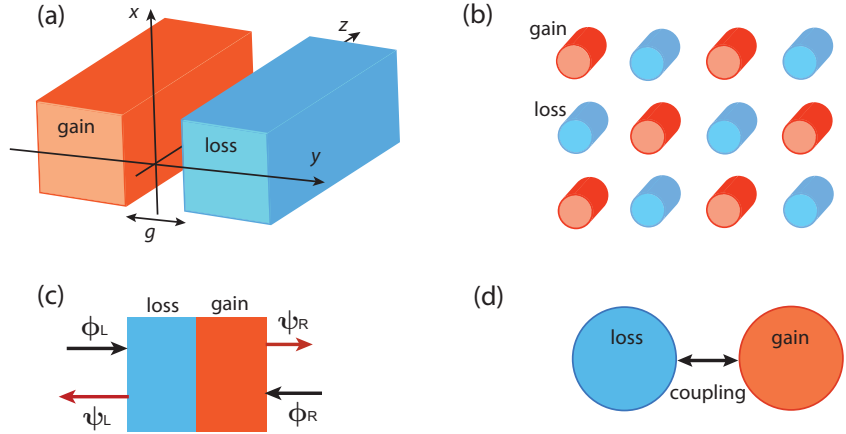}
%
%
\caption{ Schematics of state-of-art PT-symmetric structures: (a) coupled waveguides, (b) two-dimensional photonic lattices, (c) multilayer systems and (d) microresonators.
}
\label{fig:Sect3}       
\end{figure}

Coupled waveguides were the first candidates for observation of the parity-time symmetry.
In~\cite{ref:ruter2010} it was fabricated a gradient-index system with balanced loss and gain, the gain being guaranteed by the photorefractive nonlinearity of Fe-doped LiNbO$_3$. The detailed description of the physics of coupled waveguides has been provided in the previous section. Here we discuss a gain-free technique for observation of the PT symmetry demonstrated in practice in Ref.~\cite{ref:guo2009}.
The idea behind the passive PT symmetry is to carry out transformation of the fields in a purely lossy system with the aim of reducing the governing equation to that describing PT-symmetric systems.
In fact, using the gauge transformation $\ket{\psi}=\exp(-\gamma z)\ket{\tilde\psi}$,
we can rewrite Eq.~\eqref{eq:schroedinger_opt} of the passive system as 
\begin{equation}
 \ii\pdv{z}\ket{\tilde{\psi}}=\hat{\tilde{H}}\ket{\tilde{\psi}}
 \label{eq:paraxEq1}
\end{equation}
Now we claim that Eq.~\eqref{eq:paraxEq1} describes a PT-symmetric system with the effective Hamiltonian $\hat{\tilde{H}}$ which corresponds to the system with the effective permittivity
\begin{equation}
  \tilde{\varepsilon}(x,y)=\varepsilon(x,y)+2\ii\frac{n_0\gamma}{k_0}.
  \label{eq:eps_eff}
\end{equation}
$\tilde{\varepsilon}$ satisfies condition $\tilde{\varepsilon}(x,y)=\tilde{\varepsilon}^\ast(-x,-y)$ and the permittivity of the passive system meets
\begin{equation}
   \varepsilon(x,y)=\varepsilon^\ast(-x,-y)-4\ii\frac{n_0\gamma}{k_0}.
\end{equation}
Assuming that one of the waveguides is transparent (lossless) with $\varepsilon_g = \varepsilon_g^\ast$, one can easily determine the permittivity of the lossy waveguide as $\varepsilon_l = \varepsilon_g - 4 \ii n_0\gamma/k_0$.
In spite of the gain-free system does not have a true PT symmetry, it still possesses some features inherent in PT-symmetric systems as PT-symmetry breaking observed in Ref.~\cite{ref:guo2009}.
Passive PT symmetry is a smart technique to ease validation of PT-symmetry effects and its applicability.

A number of surprising effects arise in guiding systems under gain-loss modulation in a PT-symmetric manner.
The modulation shifts positions of exceptional points resulting in the directional amplification, when the phase transition is made with a threshold in one direction, being thresholdless in the opposite direction \cite{ref:song2019}.
Similarly, on the boundary between the metallic substrate and PT-symmetrically modulated dielectric slab, there is a unidirectional excitation of the surface plasmon-polaritons \cite{ref:wang2017}.
In a waveguide as an isotropic slab sandwiched between oppositely biased gyrotropic media, topologically protected guided modes arise.
PT symmetry in such a system introduces exceptional points, where electromagnetic modes are slow-light and linearly-growing \cite{ref:gangaraj2018}.
Slow light phenomenon is known to be associated with degeneracy of the modes (matching of their propagation constants).
In PT-symmetric systems, the degeneration is realized at exceptional points of mode coalescence \cite{ref:goldzak2018}.

PT-symmetric laser waveguide was fabricated in \cite{ref:yao2019}.
Gain and loss are electrically controlled to achieve a lasing threshold in the range of PT-symmetry violation.
By interplay of two guided modes there were distinguished several phases, the lasing within which being confirmed experimentally.

Two coupled waveguides experience optical forces originating from excited eigenmodes.
The forces qualitatively change at exceptional points and may result in pushing and pulling of one waveguide to another.
When the forces deflect the waveguides, they may induce the phase transition through changing a gap between them \cite{ref:xu2018}.
An unusual power flow in the PT-symmetric coupled waveguide results in an off-diagonal stress tensor components causing the shear along the mode propagation direction \cite{ref:miri2019}.

\subsection{Two-dimensional photonic waveguide lattices}

An array of parallel waveguides can be arranged in nodes of a lattice as demonstrated in Fig. \ref{fig:Sect3}(b).
Such a two-dimensional photonic crystal is a natural generalization of a pair of coupled waveguides.
To engage a PT symmetry in the lattice, the gain and lossy waveguides should be disposed periodically.
As other PT-symmetric systems, the lattice has exceptional points bordering phases of broken and unbroken PT-symmetric states.
At the same time, light beams propagating in lattices demonstrate beam splitting, power oscillations, nonreciprocity and secondary emissions~\cite{ref:makris2008}.
These diffraction properties are explained by nonorthogonality of the Floquet-Bloch modes of the periodic structure.

In the system of PT-symmetric periodically arranged cylinders situated at the interface between two semi-infinite media, unidirectional transmission without reflection can be achieved. It was investigated in Ref. \cite{ref:yuan2019} using the perturbation and scattering matrix theories.
Photonic graphene lattice of waveguides described using the coupled-mode techniques can be represented as the Dirac equation. The PT symmetry in such a system requires the corresponding Hamiltonian is non-Hermitian exhibiting unbroken and broken PT-symmetry phases. These theoretical findings are well confirmed in experiments \cite{ref:kremer2019}. 
In Ref. \cite{ref:weimann2017} photonic lattices were used for the proof of existence of topological interface states on a defect waveguide between two PT-symmetric media. The results propose a way of excitation of topologically protected localized states.
A PT-symmetric photonic crystal can be also designed as a group of gain cylinders paired with a group of lossy cylinders. Then surface electromagnetic waves emerge at the gain-loss interface, while exceptional points can be tuned to coalesce forming higher-order exceptional points \cite{ref:cui2019}.

\subsection{Multilayer structures}

To illustrate the basic principles of the PT symmetry, a simple multilayer system may be harnessed. The PT symmetry in multilayer structures is introduced in direction of the wave propagation, and multilayers as open systems can be described by a scattering matrix. The scattering matrix connects the input and output channels. For the multilayer system there are two input channels and two output channels as demonstrated in Fig. \ref{fig:Sect3}(c). The output fields $\psi_R = t \phi_L + r_{R} \phi_R$ and $\psi_L = t \phi_R + r_{L} \phi_L$ can be arranged as
\begin{equation}
    \left( \begin{array}{c} \psi_R \\ \psi_L \end{array} \right) = \left( \begin{array}{cc} r_R & t \\ t & r_L \end{array} \right) \left( \begin{array}{c} \phi_R \\ \phi_L \end{array} \right) \quad {\rm or} \quad 
    \left( \begin{array}{c} \psi_L \\ \psi_R \end{array} \right) = \left( \begin{array}{cc} t & r_L \\ r_R & t \end{array} \right) \left( \begin{array}{c} \phi_R \\ \phi_L \end{array} \right), \label{outSin}
\end{equation}
where $\phi_{L,R}$ are the input fields, $t$ is the transmission coefficent and $r_L$ and $r_R$ are the reflectrion coefficients to the left and to the right sides, respectively. Transmission coefficient $t$ does not depend on the direction of wave incidence owing to reciprocity of the system. 

Equation (\ref{outSin}) shows that the scattering matrix as a matrix between input and output fields can be defined in two different ways by means of the permutation of the output channels. Such a nominal designation is expected to be unimportant. However, since the scattering matrix eigenvalues are different for two matrices in Eq. (\ref{outSin}), but, as generally accepted, should predict exceptional points, a dilemma arises, which scattering matrix is appropriate \cite{ref:longhi2010,ref:ge2012}. 
The problem of uniqueness of the scattering matrix of the PT-symmetric system has been successfully solved in Ref.~\cite{ref:novitsky2020} using the direct connection of the scattering matrix $\hat S$ with the PT-symmetric Hamiltonian $\hat H$ of the one-dimensional multilayer system as $\hat S = \exp(\ii\hat H (t-t_0)/\hbar)$.
Correct positions of the exceptional points then read as $s_{1,2} = t \pm \sqrt{r_L r_R}$, where the scattering matrix defined by the right-hand equation in Eq. (\ref{outSin}) is employed.
Exceptional points of the scattering matrix with permuted channels given by the another scattering matrix approximate the lasing onset.

PT-symmetric multilayer systems are widely studied with the aim of enhancement of physical phenomena near exceptional points.
A PT-symmetric bilayer exhibits a giant Goos-H\"anhen shift at specific angles of incidence \cite{ref:cao2019}.
The enhancement is explained by excitation of surface modes at the interface between the gain and lossy slabs facilitating the quasi-BIC (quasi-bound state in the continuum) states.
An ordinarily weak spin-orbit interaction of light can be also significantly intensified in vicinity of exceptional points of the PT-symmetric bilayer.
Interaction of light spin and orbital momenta is coined as the spin Hall or Imbert-Fedorov effect and results in a lateral shift of a light beam.
Near exceptional points of the PT-symmetric bilayer the reflection coefficient experiences negligible values and abrupt phase shift enhancing the lateral beam displacement, though it takes zero value at the exceptional point \cite{ref:zhou2019}.
Graphene sheets in PT-symmetric multilayer systems can be used for modulation of an exceptional point position via tuning their surface conductivity \cite{ref:chen2019}. Light also makes a bilayer to move. Optical forces exerting on PT-symmetric multilayer structures can be both pushing and pulling depending on the direction of light and realization of the broken or unbroken PT-symmetric phase \cite{ref:alaee2018}.

A PT-symmetric multilayer structure can be used as a laser.
More precisely, the laser must be simultaneously a coherent perfect absorber~\cite{ref:chong2011,ref:wong2016}.
In the broken PT-symmetry phase, an illumination from one side is normally amplified, while a coherent illumination from both sides of the multilayer system is absorbed owing to interference.
Laser-absorber modes arise, when a pole and a zero of a scattering matrix approach each other on the real axis in the plane of complex frequency.
Finite-difference-time-domain (FDTD) calculations basically confirm predictions of the transfer-matrix and scattering-matrix approaches and show an enormous increase in the output intensity, when the laser threshold is achieved \cite{ref:novitsky2020}.
If the pole is not on the real axis, the lasing onset is still feasible at a greater threshold.
In realistic PT-symmetric systems, realization of the gain occurs in quantum systems and requires accounting for the saturation effect.
This means that the PT-symmetric system has to be nonlinear.
In Ref.~\cite{ref:novitsky2018light} it was considered a two-level resonant medium described by the Maxwell-Bloch equations.
Due to the saturation, the condition for PT symmetry is approximately valid and the system can be named as a non-Hermitian one \cite{ref:bartoniii2018}.
Saturation may result in novel effects, e.g., locking of the light propagation by the PT-symmetry breaking \cite{ref:novitsky2018light}.
Steady-state solutions for PT-symmetric multilayer structures with nonlinear refractive indices of gain and loss media are investigated in Ref.~\cite{ref:witonski2017} using a modified transfer-matrix method.
The bistable behavior of the transmitted and reflected intensities was studied together with unidirectional invisibility and coherent perfect absorption versus the input and saturation intensities.

\subsection{Microresonators}

PT symmetry can be realized on a resonator platform promising for interesting applications on a chip. In Fig. \ref{fig:Sect3}(d) we show a typical system comprising two coupled gain and loss cavities.
It is shown in Ref.~\cite{ref:zhong2019} that a non-Hermitian optical microring resonator coupled to a waveguide can be used as an asymmetric absorber, if a mirror is placed on one side of the waveguide.
Robustness of the asymmetric absorption is explained by the emergence of the chiral exceptional surface, which can be also exploited for directional absorption control.
Unidirectional lasing and coherent perfect absorption can be achieved using unidirectional destructive interferences being realized with an asymmetrically coupled passive resonator chain \cite{ref:jin2018}.
Asymmetry in coupling breaks the reciprocity in transmission due to the destructive interference.
A PT-symmetric side-coupled resonator can be realized using unidirectional lasing together with perfectly absorbing resonators and can result in simultaneous unidirectional lasing and perfect absorption effects.
In a similar fashion, the spectral singularities of scattering matrix can be investigated in a PT-symmetric two-arm Aharonov-Bohm interferometer~\cite{ref:jin2018}.
Such spectral singularities can be tailored to realize symmetric, asymmetric and unidirectional lasing onsets.

Non-Hermitian resonators are widely used as extremely sensible sensors at exceptional points. The sensitivity can be even more increased for higher-order exceptional points, at which more than two eigenvalues of a non-Hermitian Hamiltonian coincide. In this case, the frequency splitting stems from a perturbation $\varepsilon \ll 1$ follows the law $\varepsilon^{1/N}$, where $N$ is the order of the exceptional point \cite{ref:wiersig2014}. Since the susceptibility $d(\varepsilon^{1/N})/d\varepsilon$ diverges at $\varepsilon = 0$, the sensitivity can be arbitrary high. This idea was experimentally validated in a PT-symmetric ternary (loss-neutral-gain) micro-ring system \cite{ref:hodaei2017} and micro-toroid cavity
\cite{ref:chen2017}, the non-Hermiticity of the latter being introduced by a scatterer resulting in coupling eigenmodes of the cavity.
Sensitivity of parameter estimation can be analyzed using the formalism of quantum Fisher information without referring to a specific measurement scheme. The average of all merging eigenstates cancels out the divergence at the singularity resulting to a finite value at the exceptional point \cite{ref:chen2019}.

PT-symmetric \cite{ref:ren2017} and anti-PT-symmetric \cite{ref:carlo2019} optical gyroscopes were proposed on the basis of ring resonators coupled to a waveguide. Frequency splitting and, hence, sensitivity in gyroscopes are normally limited by the ring dimensions. In PT-symmetric gyroscopes, the frequency splitting is independent of the ring radius and, therefore, the phase shift of interference fringes is enhanced in vicinity of exceptional points.
A unique ``superluminal'' lasing may be used as a sensor and can be obtained in a broken PT-symmetry phase of the white-light cavity that consists of gain and lossy coupled micro-resonators \cite{ref:scheuer2018}.

Finally, PT-symmetric microcavities possess indispensable nanophotonic properties for suppression of spontaneous relaxation rate \cite{ref:akbarzadeh2019}. In the next section we will discuss this subject in detail.

\section{Purcell Effect in PT-symmetric waveguides}
\label{sec:purcel_pt_wg}
In 1946, E. M. Purcell predicted that the spontaneous emission rate of a light source is not
solely an intrinsic property of the source but is affected by the optical environment~\cite{ref:purcell1946}.
This effect is now referred to as Purcell effect.
The Purcell factor is defined as
\begin{equation}
  F_p=\frac{P_\mathrm{system}}{P_0},
\end{equation}
where $P_\mathrm{system}$ denotes the power of an emitter radiated into a particular optical system
and $P_0$ is the power of the same emitter radiated into vacuum or free space.
Purcell factor is a common figure of merit to describe the emission enhancement induced by feedback of the source with a particular optical system.
Alternatively, Purcell factor can be defined in terms of spontaneous emission rate
\begin{equation}
  F_p=\frac{\tau_0}{\tau_\mathrm{system}},
\end{equation}
where $\tau_0$ is the spontaneous emission lifetime in vacuum and $\tau_\mathrm{system}$ is the lifetime of the emitter in the particular system of interest.

The interaction between the emitter and its environment is formally described by Fermi's golden rule which states that the probability for spontaneous emission is proportional to the (photonic) local density of states (LDOS).
LDOS, in turn, is proportional to the imaginary part of Green's dyadic $\hat{G}$ at the position of the emitter~\cite{ref:novotny2012}
\begin{equation}
  \rho_p(\vb{r}_0,\omega)=\frac{6\omega}{\pi c^2}
  \left[\vu{p}\Im\hat{G}(\vb{r}_0,\vb{r}_0;\omega)\vu{p}\right],
\end{equation}
where $\vb{r}_0$ is emitter position and $\vu{p}$ denotes unit vector of the dipole orientation.

It is well known since the Purcell's work~\cite{ref:purcell1946} that the strong Purcell enhancement occurs in resonant systems where the light is confined to small volumes.
More recent work suggests that giant enhancements can occur via the less familiar
Petermann effect~\cite{ref:petermann1979,ref:siegman1989,ref:yoo2011}.
The Petermann enhancement factor is a measure of non-orthogonality
of the modes in non-Hermitian systems and it appears to diverge when two modes coalesce at an exceptional point (EP).
In the work of Pick~et~al~\cite{ref:pick2017} authors develop a general theory of the spontaneous emission at exceptional points.
They show that traditional theories of spontaneous emission fail in case of degenerate resonances occurring at EPs and lead to infinite Purcell factors.
Approach presented in~\cite{ref:pick2017} is based on the perturbation theory which properly accounts for degeneracies at EPs by using Jordan vectors.
Within this approach authors prove that actual enhancement factors are finite, but can still be significant (about hundreds) in gain-aided and higher-order EP systems.

Interestingly, that not only enhancement but rather suppression of spontaneous decay rate can occur in PT-symmetric systems.
Akbarzadeh~et~al~in~Ref.~\cite{ref:akbarzadeh2019} show that a PT-symmetric planar cavity is able to suppress the spontaneous relaxation rate of a two-level atom below the vacuum level.
Recent work of Khanbekyan~and~Wiersig reports on decay suppression of spontaneous emission of a single emitter in a high-$Q$ cavity at exceptional points~\cite{ref:khanbekyan2020}.

The Purcell factor can be calculated separately for each of the discrete scatter-
ing channels.
For instance, just a couple of years ago,
the Purcell effect in the mode of the basic element of PIC planar waveguide was introduced within the scattering matrix formalism~\cite{ref:ivanov2017}.

It has recently been shown in the context of single molecule detections that the power emitted from a molecule into a single mode fiber can be elegantly calculated using the reciprocity theorem of electromagnetic theory~\cite{ref:then2014}.
In the work~\cite{ref:schulz2018} authors propose a reciprocity approach to calculate the emission enhancement for emitters coupled to arbitrary resonant or non-resonant open optical systems.
They calculate the \emph{modal Purcell factor} --- the quantity which measures the power emitted by an emitter situated in the vicinity of a device into a particular propagating mode normalized by the power radiated by the same emitter into the free space.

\subsection{Reciprocity approach}
In this section, we generalize the reciprocity approach formulated in~\cite{ref:schulz2018} to the case when the propagating modes are not orthogonal. We probe the method by calculation of the modal Purcell factor in PT-symmetric coupled waveguide system.
In the following section we will obtain closed-form expressions for modal Purcell factor describing the system in terms of coupled modes.

We consider a current source (current density distribution $\vb{J}_1$) situated in the vicinity of some optical system with two exit ports at $z_1$ and $z_n$ ~\cite{ref:schulz2018}.
For brevity, we use optical state-vector notation for 4-component vector joining transverse electric and magnetic fields introduced in Eq.~\eqref{eq:state_vector}.
In this way we can describe the fields of guiding (and leaking) modes.
For the $i$th mode we write
\begin{equation}
  \ket{M_i(z)}=\mqty(\vb{E}_{t,i}(x, y, z) \\ \vb{H}_{t,i}(x, y, z))=\ket{i}\ee^{-\ii\beta_iz},
  \label{eq:Mode_i}
\end{equation}
where
\begin{equation}
  \ket{i}=\mqty(\vb{e}_{t,i}(x, y) \\ \vb{h}_{t,i}(x, y))
  \label{eq:mode_i}
\end{equation}
and
\begin{equation}
  \mqty(\vb{E}_{t,i}(x, y, z) \\ \vb{H}_{t,i}(x, y, z)) =
  \mqty(\vb{e}_{t,i}(x, y) \\ \vb{h}_{t,i}(x, y))\ee^{-\ii\beta_iz}.
\end{equation}

The fields excited by the current source $\vb{J}_1$ at the cross-section of exit ports can be expanded into a set of modes as follows
\begin{align}
  \ket{\psi_1(z_1)}&=\sum_{i} A_{i,z_1} \ket{i,z_1}, \nonumber \\
  \ket{\psi_1(z_n)}&=\sum_{i} A_{-i,z_n} \ket{-i,z_n}.
\label{eq:expansion}
\end{align}
Here $A_{i,z_1}$ and $A_{-i,z_n}$ are the amplitudes of the modes propagating forward to port $z_1$ and backward to port $z_n$, respectively,
$\ket{i,z_1}$, $\ket{-i,z_n}$ are respectively eigenmodes of ports $z_1$ and $z_n$ propagating from the cavity.

In our notations the Lorentz reciprocity theorem
\begin{equation}
  \intop_{\delta V} \left( \vb{E}_{1}\times\vb{H}_{2} - \vb{E}_{2}\times\vb{H}_{1} \right)
  \cdot \vu{z} \dd x \dd y
  = \intop_{V} \left(\vb{E}_{2} \cdot \vb{J}_{1} - \vb{E}_{1} \cdot \vb{J}_{2} \right) \dd V.
  \label{eq:reciprocity}
\end{equation}
should be rewritten as
\begin{multline}
\braket{\psi_1(z_1)}{\psi_2(z_1)}-\braket{\psi_2(z_1)}{\psi_1(z_1)}
-\braket{\psi_1(z_n)}{\psi_2(z_n)}+\braket{\psi_2(z_n)}{\psi_1(z_n)}\\
  = \intop_{V}\left( \vb{E}_{2}\cdot\vb{J}_{1}-\vb{E}_{1}\cdot\vb{J}_{2} \right)\dd V,
  \label{eq:reciprocity1}
\end{multline}
where $\delta V$ is the surface enclosing the cavity volume $V$ between two planes $z=z_1$ and $z=z_n$.
In Eq.~\eqref{eq:reciprocity1}, ${\bf J}_1$ and $\ket{\psi_1}$ are defined above, while the source ${\bf J}_2$ and the fields $\ket{\psi_2}$ produced by it can be chosen as we need.
Let the source current ${\bf J}_2$, being outside the volume $V$ (${\bf J}_2 = 0$), excite a single mode $\ket{-k,z_1}$.
In general, this mode is scattered by the cavity $V$ and creates the set of transmitted and reflected modes as discussed in \cite{ref:schulz2018}:

\begin{align}
  \ket{\psi_{2}(z_1)} =& B_{-k, z_1} \ket{-k,z_1}
  +\sum_i B_{i, z_1} \ket{i,z_1},
  \label{eq:psi2z1}
  \\
  \ket{\psi_{2}(z_n)} =& \sum_i B_{-i,z_n}\ket{-i,z_n}. 
  \label{eq:psi2zn}
\end{align}

Using the orthogonality condition~\eqref{eq:orth} and the symmetry relations~\eqref{eq:forward_backward_braket} we obtain the inner products of the fields
\begin{subequations}
  \begin{multline}
    \braket{\psi_1(z_1)}{\psi_2(z_1)} = 
    \sum_{i}A_{i,z_1}B_{-k,z_1}\bra{i,z_1}\ket{-k,z_1}
    + \sum_{i,j}A_{i,z_1}B_{j,z_1}\bra{i,z_1}\ket{j,z_1}\\
    =-2A_{k,z_1}B_{-k,z_1}N_k +
    2\sum_{i}A_{i,z_1}B_{i,z_1}N_{i,z_1}.
  \end{multline}
  \begin{multline}
    \braket{\psi_2(z_1)}{\psi_1(z_1)} =
    \sum_{i}A_{i,z_1}B_{-k,z_1}\bra{-k,z_1}\ket{i,z_1}
    +\sum_{i,j}A_{i,z_1}B_{j,z_1}\bra{i,z_1}\ket{j,z_1}\\
    =2A_{k,z_1}B_{-k,z_1}N_{k,z_1}+2\sum_{i}A_{i,z_1}B_{i,z_1}N_{i,z_1}.
  \end{multline}
  \begin{multline}
    \braket{\psi_1(z_n)}{\psi_2(z_n)}=\braket{\psi_2(z_n)}{\psi_1(z_n)}=
    \sum_{i, j}A_{-i,z_n}B_{-j,z_n}\bra{-i,z_n}\ket{-j,z_n}\\
    =2\sum_{i}A_{-i,z_n}B_{-i,z_n}N_{i,z_n},
  \end{multline}
  \label{eq:rhs_recipr}
\end{subequations}
where $N_{i,z_{1(n)}}$ the norm of the mode $\ket{i,z_{1(n)}}$ as defined in \eqref{eq:orth}.

By substituting these equations into Eq.~\eqref{eq:reciprocity1}, we arrive at the amplitude $A_{k,z_1}$ of the mode excited by the source current $\vb{J}_1$ 
\begin{equation}
  A_{k,z_1}=-\frac{1}{4B_{-k,z_1}N_{k,z_1}} 
  \intop_{V} \vb{E}_{2,-k} \cdot \vb{J}_{1} \dd V,
  \label{eq:minus4ab}
\end{equation}
where $\vb{E}_{2,-k} = B_{-k, z_1}\vb{e}_{-k}(x,y)\ee^{\ii \beta_k (z-z_1)}$ is the electric field created by the excitation of the system with reciprocal mode $\ket{-k,z_1}$ at the port $z_1$.

As an emitter we consider a point dipole oscillating at the circular frequency $\omega$ and having the current density distribution
\begin{equation}
  \vb{J}_{1}\left( \vb{r} \right) =
  \ii\omega\vb{p}\delta\left( \vb{r}-\vb{r}_0 \right),
  \label{eq:dipole_current}
\end{equation}
where $\vb{p}$ is the dipole moment of the emitter and $\vb{r}_0$ is its position.
Then we are able to carry out the integration in Eq.~\eqref{eq:minus4ab} and obtain
\begin{equation}
  A_{k,z_1}=-\frac{\ii\omega}{4 B_{-k, z_1}N_{k,z_1}} \vb{E}_{2,-k}
  \left( \vb{r}_0 \right) \cdot \vb{p}.
  \label{eq:Akz1}
\end{equation}
Here we observe a dramatic difference compared to the Hermitian case considered in Ref.~\cite{ref:schulz2018}.
This difference appears due to the fact that now the expansion coefficients $A_{k, z_1}$ are not directly related to the powers carried by the modes.
Finding a power carried by a specific mode is a challenge.
To circumvent this challenge, we propose a calculation of the total power carried by the set of modes as we describe below.

The power emitted by the current source $\vb{J}_1$ into the port $z_1$ can be written as
\begin{equation}
  P = \frac{1}{2} \mathrm{Re}\intop_{z=z_1} \left( \vb{E}_{1} \times \vb{H}^{*}_{1} \right) \cdot
  \vu{z} \dd x \dd y = \frac{1}{2}\Re\braket{\psi_1(z_1)}{\psi_1^\ast(z_1)}.
  \label{eq:power1}
\end{equation}
Expanding the electromagnetic fields $\ket{\psi_1(z_1)}$ according to Eq.~(\ref{eq:expansion})
we represent the power transmitted through the port Eq. (\ref{eq:power1}) as follows
\begin{equation}
  P = \mathrm{Re}\sum_{k,l}A_{k, z_1} A^{*}_{l, z_1} P_{kl},
  \label{eq:power1_expand}
\end{equation}
where $P_{kl}$ is the so called cross-power equal to the Hermitian inner product of the modal fields
\begin{equation}
  P_{kl,z_1} = \frac{1}{2} \braket{k,z_1}{l^*,z_1} =
  \frac{1}{2}\intop_{z=z1}\left( \vb{e}_{k,z_1} \times \vb{h}^{*}_{l,z_1} \right)\vdot\vu{z}\dx\dy.
  \label{eq:cross_power}
\end{equation}
For $k = l$ the cross-power reduces to the mode power $P_k = \Re P_{kk}$. By considering the expansion coefficients \eqref{eq:Akz1} we rewrite the power (\ref{eq:power1_expand}) in terms of the reciprocal fields $\vb{E}_{2,-k}$ as
\begin{multline}
  P = \frac{\omega^{2}}{16}
    \Re
    \sum_{k,l}
    \frac{
  ( \vb{E}_{2,-k}\left( \vb{r}_0 \right)\cdot \vb{p} )
  ( \vb{E}_{2,-l}^{*}\left( \vb{r}_0 \right)\cdot \vb{p}^\ast )
    }
    {B_{-k}B_{-l}^{*} N_{k} N_{l}^*}
    P_{kl}
    \\
  =\frac{\omega^{2}}{16}
    \Re
    \sum_{k,l}
    \frac{
  (\vb{e}_{-k}\left( x_0, y_0 \right)\cdot \vb{p} )
  ( \vb{e}_{-l}^{*}\left( x_0, y_0 \right)\cdot \vb{p}^\ast )
    }
    {N_{k} N_{l}^*}
    P_{kl}.
    \label{eq:power_final}
\end{multline}
The last equality is the consequence of the substitution of $\vb{E}_{2,-k}$ at the emitter position ${\bf r}_0 = (x_0, y_0, z_0)$ considering the negligible dimensions of the cavity $z_1\approx z_n\approx z_0$.
Note that here we dropped $z_1$ subscripts.

To find the Purcell factor we divide Eq. (\ref{eq:power_final}) by the power emitted by the same dipole into the free space
\begin{equation}
  P_{0} = \frac{\mu_{0}}{12\pi c} \omega^{4} |p|^{2},
  \label{eq:P0}
\end{equation}
where $\mu_0$ is the vacuum permeability and $c$ is the speed of light in vacuum.
The dipole moment, located in the $xy$ plane, can be presented using the unit vector $\hat {\bf p}$ as follows
\begin{equation}
 \vb{p}=p\vu{p},
\end{equation}
therefore,
\begin{equation}
  \vb{E}_{2,-k}(\vb{r}_0)\cdot\vb{p}=\vb{E}_{2,-k}(\vb{r}_0)\cdot\vu{p}p=E_{p,k}(\vb{r}_0)p.
  \label{eq:Edotp}
\end{equation}
Here $E_{p,k}$ denotes projection of the vector $\vb{E}_{2,-k}$ onto the dipole orientation vector $\vu{p}$
\begin{equation}
  E_{p,k}=\vb{E}_{2,-k}\cdot\vu{p}.
  \label{eq:cos_alpha}
\end{equation}
Then the Purcell factor reads
\begin{equation}
  F_{p} = \frac{P}{P_0}
  =\frac{3\pi c}{4\omega^{2} \mu_{0}} \mathrm{Re}
  \sum_{k,l}
    \frac{
      e_{p,k}\left( x_0, y_0 \right)
      e_{p,l}^{*}\left( x_0, y_0 \right)
    }{N_{k} N_{l}^*}P_{kl}.
  \label{eq:Fpurcell}
\end{equation}

It is convenient to rewrite Eq. \eqref{eq:Fpurcell} through the normalized fields as
\begin{equation}
  F_{p}=
  \frac{3\pi c}{4\omega^{2} \mu_{0}}\sum_{kl}\hat{e}_{p,k}\hat{e}^*_{p,l}K_{kl}\hat{P}_{kl},
  \label{eq:Fp_normed}
\end{equation}
where we have introduced power-normalized modal electric fields
\begin{equation}
  \hat{\vb{e}}_{2, i} = \frac{\vb{e}_{2,i}}{\sqrt{P_{i}}}
  \label{eq:e_power_normalized}
\end{equation}
and normalized cross-power coefficients
\begin{equation}
  \hat{P}_{kl} = \frac{1}{\sqrt{P_kP_l}}P_{kl}.
\end{equation}
Here we generalize the Petermann factor defined in Section~\ref{subsec:petermann}
\begin{equation}
  K_i = K_{ii}
  \label{eq:Petermann_factor}
\end{equation}
defining cross-mode Petermann factor
\begin{equation}
  K_{kl} =
  \frac{\braket{k}{k^*}}{\braket{k}}\frac{\braket{l}{l^*}^*}{\braket{l}^*}.
  \label{eq:cross_Petermann_factor}
\end{equation}
The modal Purcell factor can be naturally divided into two parts, the first of which is the sum of all diagonal $(k=l)$ terms, while the second part is the sum of off-diagonal $(k\neq l)$ terms:
\begin{equation}
  F_p = F_{p, \mathrm{diag}} + F_{p,\mathrm{off-diag}} =
  \sum_kF_{p, k}+\sum_{k\neq l}F_{p, kl},
  \label{eq:Fp_two_terms}
\end{equation}
where
\begin{align}
  F_{p, i}&=\frac{3\pi c}{4\omega^{2} \mu_{0}}
    \abs{\hat{e}_{p,k}}^2 K_{i},
  \label{eq:Fp_diag}\\
  F_{p, kl}&=\frac{3\pi c}{4\omega^{2} \mu_{0}}
  \hat{e}_{p,k}\hat{e}^*_{p,l} K_{kl}\hat{P}_{kl}.
  \label{eq:Fp_off-diag}
\end{align}
In the Hermitian case, the off-diagonal terms \eqref{eq:Fp_off-diag} reduce to zero due to the regular orthogonality of the modes expressed by $\hat{P}_{kl}=\delta_{kl}$.
That is why the Purcell factor \eqref{eq:Fp_normed} applied to Hermitian systems coincides with the expression in Ref.~\cite{ref:schulz2018}.

\subsection{Modal Purcell factor within the Coupled Mode Theory}
\label{subsec:cmt}

\paragraph{\textbf{PT-symmetric regime}}
To find the modal Purcell factor for the coupled waveguide system in PT-symmetric regime we substitute
the modes in the form \eqref{eq:eigmodes_sym} into expression \eqref{eq:Fp_normed}.

One more assumption is introduced for the sake of simplicity:
\begin{equation}
  \braket{g}{g^*}=\braket{l}{l^*}=1.
\end{equation}
It implies that the Hermitian norms of the isolated modes
are equal to the non-Hermitian norms or, in other words,
the Petermann factors for the modes equal unity.

Then the quantities $K_{kl}$ and $\hat{P}_{kl}$ can be written in the closed form as
\begin{subequations}
\begin{align}
  K_{1}=\frac{\abs{\braket{1}{1^*}}^2}{\abs{\braket{1}}^2}
  &=\frac{2}{1+\cos2\theta},\\
  K_{2}=\frac{\abs{\braket{2}{2^*}}^2}{\abs{\braket{2}}^2}
  &=\frac{2}{1+\cos2\theta},\\
  K_{12}=\frac{\braket{1}{1^*}}{\braket{1}}
  \frac{\braket{2}{2^*}^*}{\braket{2}^*}
  &=\frac{2(1+\ee^{-\ii2\theta})^2}{(1+\cos2\theta)^2},\\
  K_{21}=\frac{\braket{2}{2^*}}{\braket{2}}
  \frac{\braket{1}{1^*}^*}{\braket{1}^*}
  &=\frac{2(1+\ee^{+\ii2\theta})^2}{(1+\cos2\theta)^2},
  \label{eq:petermann_sym}
\end{align}
\label{eq:K_kl_sym}
\end{subequations}
\begin{subequations}
\begin{align}
  \hat{P}_{12}
  =\frac{\braket{1}{2^*}}{\sqrt{\braket{1}{1^*}\braket{2}{2^*}}}
  &=\frac{1}{2}(1-\ee^{\ii2\theta}),\\
  \hat{P}_{21}
  =\frac{\braket{2}{1^*}}{\sqrt{\braket{1}{1^*}\braket{2}{2^*}}}
  &=\frac{1}{2}(1-\ee^{-\ii2\theta}).
\end{align}
\label{eq:P_kl_sym}
\end{subequations}
Normalized field projections $\hat{e}_{p,k}$ in
the basis of isolated modes read
\begin{subequations}
\begin{align}
  \hat{e}_{p,1}&=\frac{1}{\sqrt{\frac12\bra{1}\ket{1^*}}}(\hat{e}_{p,g}+\ee^{\ii\theta}\hat{e}_{p,l})
  =\hat{e}_{p,g}+\ee^{\ii\theta}\hat{e}_{p,l},\\
  \hat{e}_{p,2}&=\frac{1}{\sqrt{\frac12\bra{2}\ket{2^*}}}(\hat{e}_{p,g}-\ee^{-\ii\theta}\hat{e}_{p,l})
  =\hat{e}_{p,g}-\ee^{-\ii\theta}\hat{e}_{p,l}.
  \label{eq:ep12_sym}
\end{align}
\end{subequations}
In above expressions $\hat{e}_{p,g}$ and $\hat{e}_{p,l}$ denote projections of the fields of
backward-propagating isolated modes onto dipole orientation.
If the emitter dipole moment is perpendicular to $\hat{z}$,
projections of backward-propagating modal fields are equal to the projections of forward-propagating ones.

Performing calculation of the modal Purcell factor \eqref{eq:Fp_normed} using
relations (\ref{eq:K_kl_sym}-\ref{eq:P_kl_sym}) we obtain
\begin{equation}
  F_{p}=F_{p,\mathrm{diag}}+F_{p,\mathrm{off-diag}}=
  \frac{6\pi c}{\omega^{2} \mu_{0}}
  (\abs{\hat{e}_{p,g}}^2+\abs{\hat{e}_{p,l}}^2).
  \label{eq:Fp_sym}
\end{equation}
Diagonal and off-diagonal terms separately take the form
\begin{subequations}
\begin{align}
  F_{p,\mathrm{diag}}&=\frac{3\pi c}{4\omega^{2} \mu_{0}}
    \frac{4}{1+\cos2\theta}(\abs{\hat{e}_{p,g}}^2+\abs{\hat{e}_{p,l}}^2),\\
  F_{p,\mathrm{off-diag}}&=-\frac{3\pi c}{4\omega^{2} \mu_{0}}
  \frac{2(1-\cos2\theta)}{1+\cos2\theta}(\abs{\hat{e}_{p,g}}^2+\abs{\hat{e}_{p,l}}^2).
\end{align}
\label{eq:Fp_terms_sym}
\end{subequations}

It is curious that although both diagonal and off-diagonal terms \eqref{eq:Fp_terms_sym}
are singular at the EP corresponding to $\theta_{EP}=\pi/2$ and $\cos2\theta_{EP}=-1$,
the singularities cancel each other making the modal Purcell factor finite and independent of $\theta$.
The modal Purcell factor (\ref{eq:Fp_sym}) depends solely on the mode profiles of the isolated modes in PT-symmetric regime.

\paragraph{\textbf{PT-symmetry-broken regime}}

To obtain the modal Purcell factor in PT-symmetry-broken regime we substitute
the modes in the form \eqref{eq:eigmodes_broken} into expression \eqref{eq:Fp_normed}.

Calculating the Petermann factors
\begin{subequations}
\begin{align}
  K_{1}&=\coth^2{\theta},\\
  K_{2}&=\coth^2{\theta},\\
  K_{12}&=-\coth^2{\theta},\\
  K_{21}&=-\coth^2{\theta},
  \label{eq:petermann_broken}
\end{align}
\end{subequations}
normalized cross-powers
\begin{subequations}
\begin{align}
  \hat{P}_{12}&=\frac{1}{\cosh\theta},\\
  \hat{P}_{21}&=\frac{1}{\cosh\theta},
\end{align}
\end{subequations}
and reciprocal modal field projections
\begin{subequations}
\begin{align}
  \hat{e}_{p,1}&=
  \frac{1}{\sqrt{\frac12(1+\ee^{-2\theta})}}(\hat{e}_{p,g}+\ii\ee^{-\theta}\hat{e}_{p,l}),\\
  \hat{e}_{p,2}&=
  \frac{1}{\sqrt{\frac12(1+\ee^{2\theta})}}(\hat{e}_{p,g}+\ii\ee^{\theta}\hat{e}_{p,l})
\end{align}
\label{eq:ep12_broken}
\end{subequations}
we straightforwardly derive the diagonal and off-diagonal terms
\begin{subequations}
\begin{multline}
  F_{p,\mathrm{diag}}=
  \frac{3\pi c}{4\omega^{2} \mu_{0}}
  \frac{2\cosh\theta}{\sinh^2\theta}
  \left((\abs{\hat{e}_{p,g}}^2+\abs{\hat{e}_{p,l}}^2)\cosh\theta-
  2\Im(\hat{e}_{p,g}^*\hat{e}_{p,l})\right),
\end{multline}
\begin{multline}
  F_{p,\mathrm{off-diag}}=
  -\frac{3\pi c}{4\omega^{2} \mu_{0}}
  \frac{2}{\sinh^2\theta}
  \left(\abs{\hat{e}_{p,g}}^2+\abs{\hat{e}_{p,l}}^2-
  2\cosh\theta\Im(\hat{e}_{p,g}^*\hat{e}_{p,l})\right)
\end{multline}
\label{eq:Fp_terms_broken}
\end{subequations}
as well as the modal Purcell factor
\begin{equation}
  F_{p}=F_{p,\mathrm{diag}}+F_{p,\mathrm{off-diag}}=
  \frac{6\pi c}{\omega^{2} \mu_{0}}
  (\abs{\hat{e}_{p,g}}^2+\abs{\hat{e}_{p,l}}^2).
  \label{eq:Fp_broken}
\end{equation}

The main result of this section is that although diagonal and off-diagonal terms
of the modal Purcell factor diverge at the EP,
the modal Purcell factor itself does not exhibit a singular behavior
when approaching to the EP either from the left or right side.

Though we do not carry out a rigorous analysis of the behavior at the EP
accounting for the degeneracy of the modes as it was done in Ref.~\cite{ref:pick2017}, the developed approach leads to the well-defined expressions \eqref{eq:Fp_sym}~and~\eqref{eq:Fp_broken} for $F_p$ at the exceptional point.

\subsection{\label{subsec:results}Numerical Example: PT-symmetric coupler}

This section presents an example of utilizing the theory developed in the previous section.
Here, we analyse an optical system consisting of two coupled rectangular waveguides with width $w$ and height $h$
separated by the distance $g$ as schematically shown in Fig.~\ref{fig:Sect3}(a).
We assume that the complex refractive indices of the left (Gain) and right (Loss) waveguides are
$n_l=n_{\rm{co}}+\ii\gamma$ and $n_r=n_{\rm{co}}-\ii\gamma$ respectively.
$n_{\rm{co}}$ is the real part of the refractive index and $\gamma>0$ is the gain/loss (non-Hermiticity) parameter.
Thus, the system of the coupled waveguides satisfies PT-symmetry condition $n(x,y)=n^*(-x,-y)$.
The refractive index of the background is assumed to be unity.

We take parameters of the waveguide coupler as $g=2$ $\mu$m,
$w=1$ $\mu$m, $h=0.5$ $\mu$m, and $n_\mathrm{co}=1.44$.
The coupler has two quasi-TE supermodes at this wavelength.
We calculated the field distribution of the guided modes of these waveguides,
shown in Figs.~\ref{fig:modes_below_ep}~and~\ref{fig:modes_above_ep}.

\begin{figure}[t!b!]
  \centering
  \includegraphics[width=0.8\linewidth]{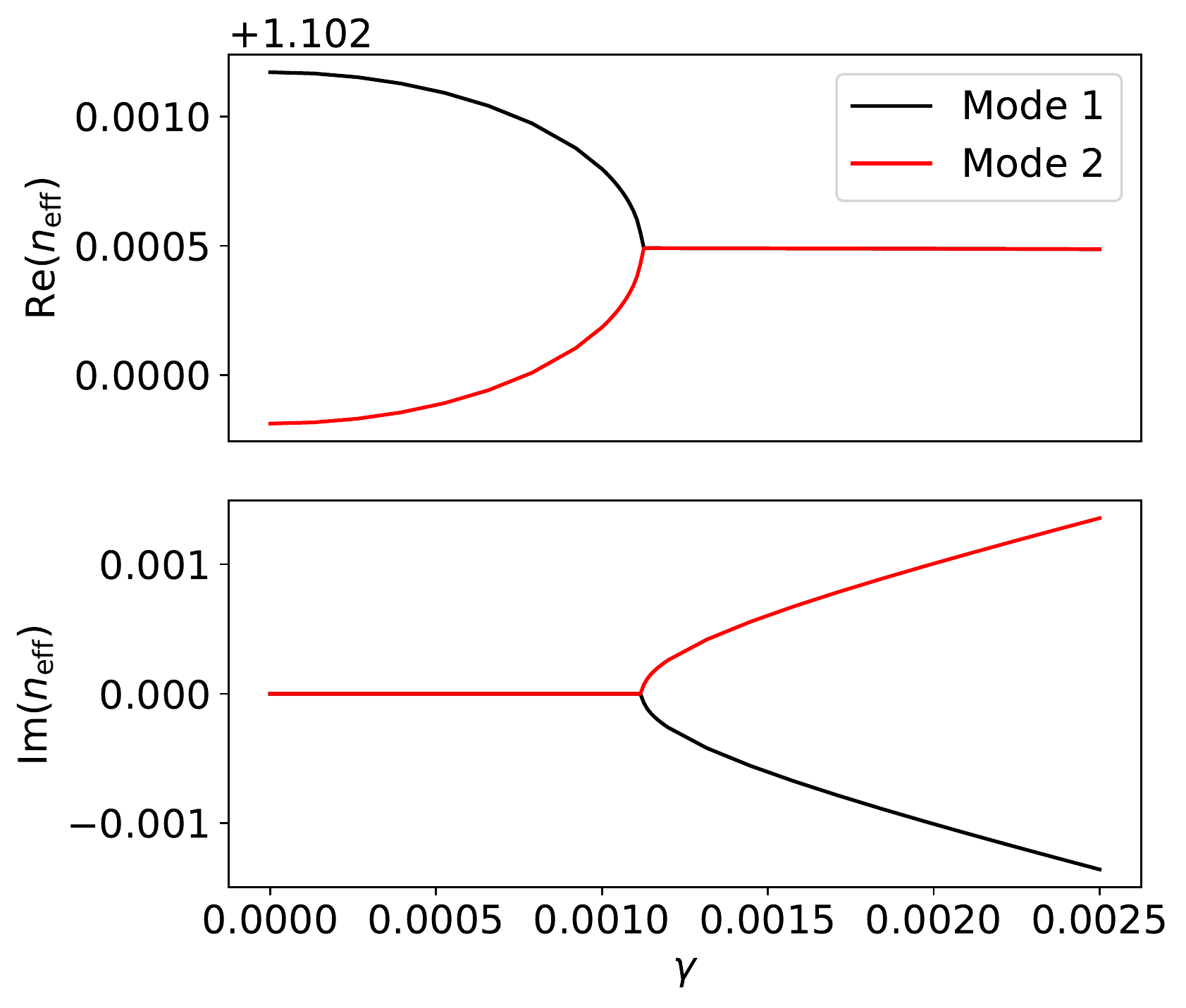}
  \caption{Effective mode indices versus the non-Hermiticity parameter $\gamma$.
  Black curves correspond to the mode $\ket{1}$.
  Red curves correspond to the mode $\ket{2}$.}
  \label{fig:neff_vs_gamma}
\end{figure}

By increasing the gain/loss parameter $\gamma$ the system passes through the regime of propagation (PT-symmetric state)
for two non-decaying supermodes to the regime of decay/amplification (PT-symmetry-broken state).
This behavior, shown by the curves in Fig.~\ref{fig:neff_vs_gamma}.

For the studied system the value of $\gamma$ corresponding to EP is $\gamma_\text{EP}=1.12\times 10^{-3}$.

\begin{figure}[t!b!]
  \centering
  \includegraphics[width=0.8\linewidth]{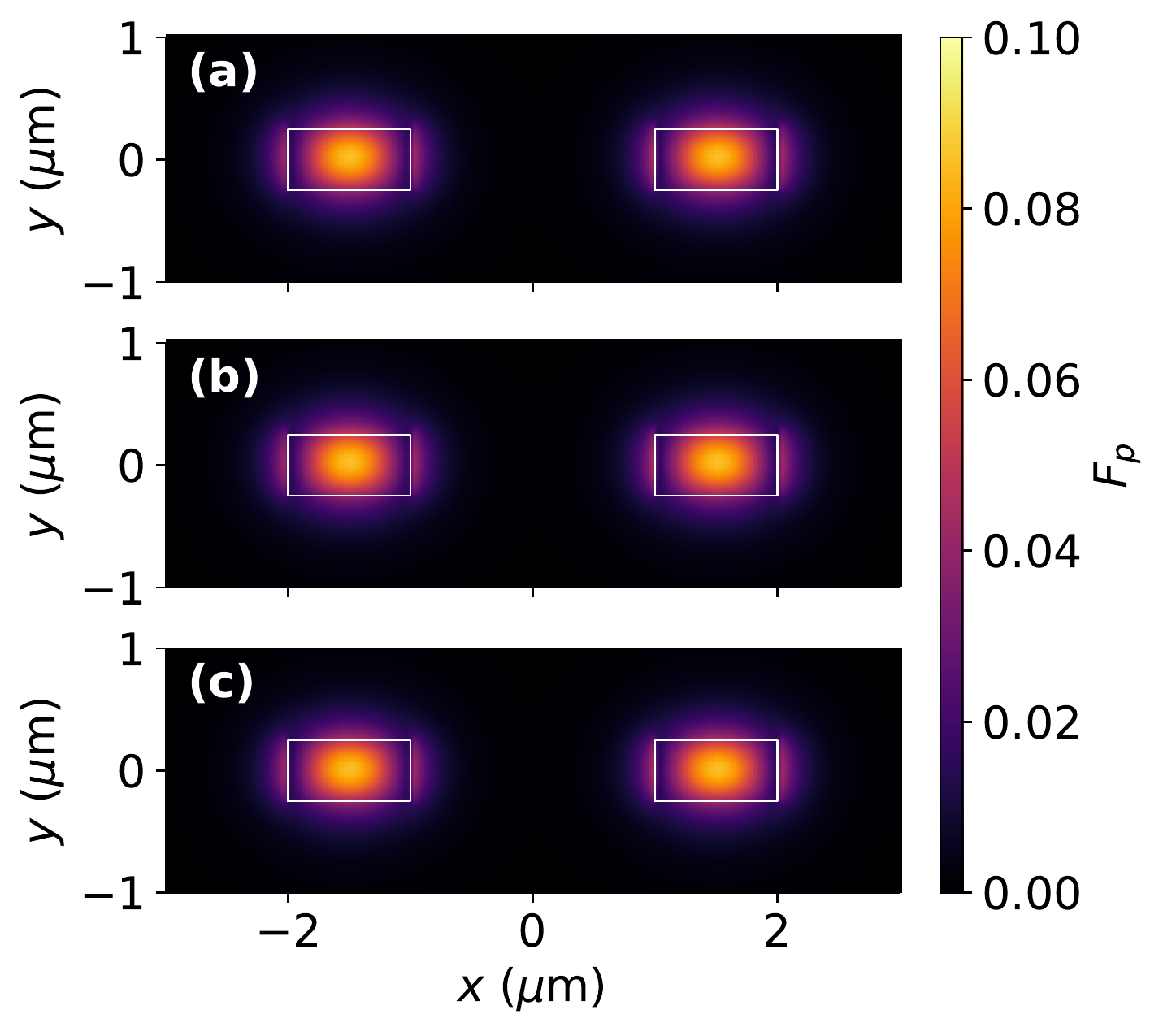}
  \caption{Purcell factor distribution in the plane ($x$, $y$) (a) for the Hermitian system characterized by $\gamma=0$, (b) in the PT-symmetric phase ($\gamma=1.02\times 10^{-3}$), (c) in the broken-PT-symmetric state ($\gamma=2.5\times 10^{-3}$).
  Parameters of the waveguide coupler: $g=2.0$ $\mu$m, $w=1$ $\mu$m, $h=0.5$ $\mu$m, and $n_\mathrm{co}=1.44$.
  }
  \label{fig:Fp_in_plane}
\end{figure}

Next, we explore the modal Purcell factor $F_p$ for the pair of quasi-TE modes.
According to Eq.~\eqref{eq:Fpurcell}, the Purcell factor is defined by the fields of the reciprocal modes at the dipole position ($x_0$, $y_0$, $z_0\approx z_1 \approx z_n$).
In Fig.~\ref{fig:Fp_in_plane}, we demonstrate the Purcell factor for an $x$-oriented dipoles as a function of
$x_0$~and~$y_0$ for different values of non-Hermiticity parameter $\gamma$.

\begin{figure}[t!b!]
  \centering
  \includegraphics[width=0.7\linewidth]{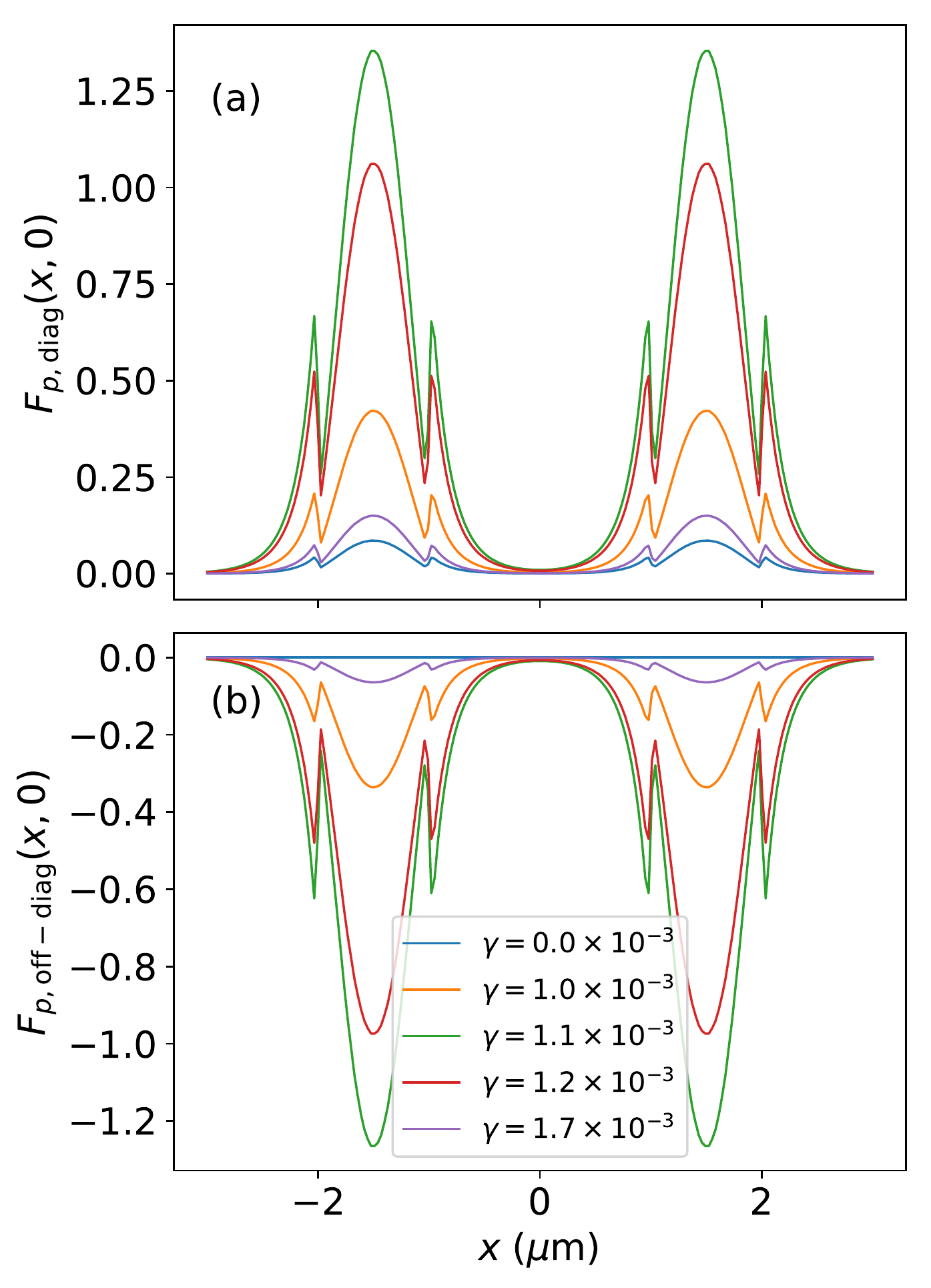}
  \caption{Distribution of the Purcell factor (a) diagonal and (b) off-diagonal terms depending
  on the emitter position $x_0$ at $y_0=0$ for different values of $\gamma$. Parameters of the coupled waveguide are given in the caption of Fig. \ref{fig:Fp_in_plane}.}
  \label{fig:Fp_gamma_y_is_0}
\end{figure}

\begin{figure}[t!b!]
  \centering
  \includegraphics[width=0.8\linewidth]{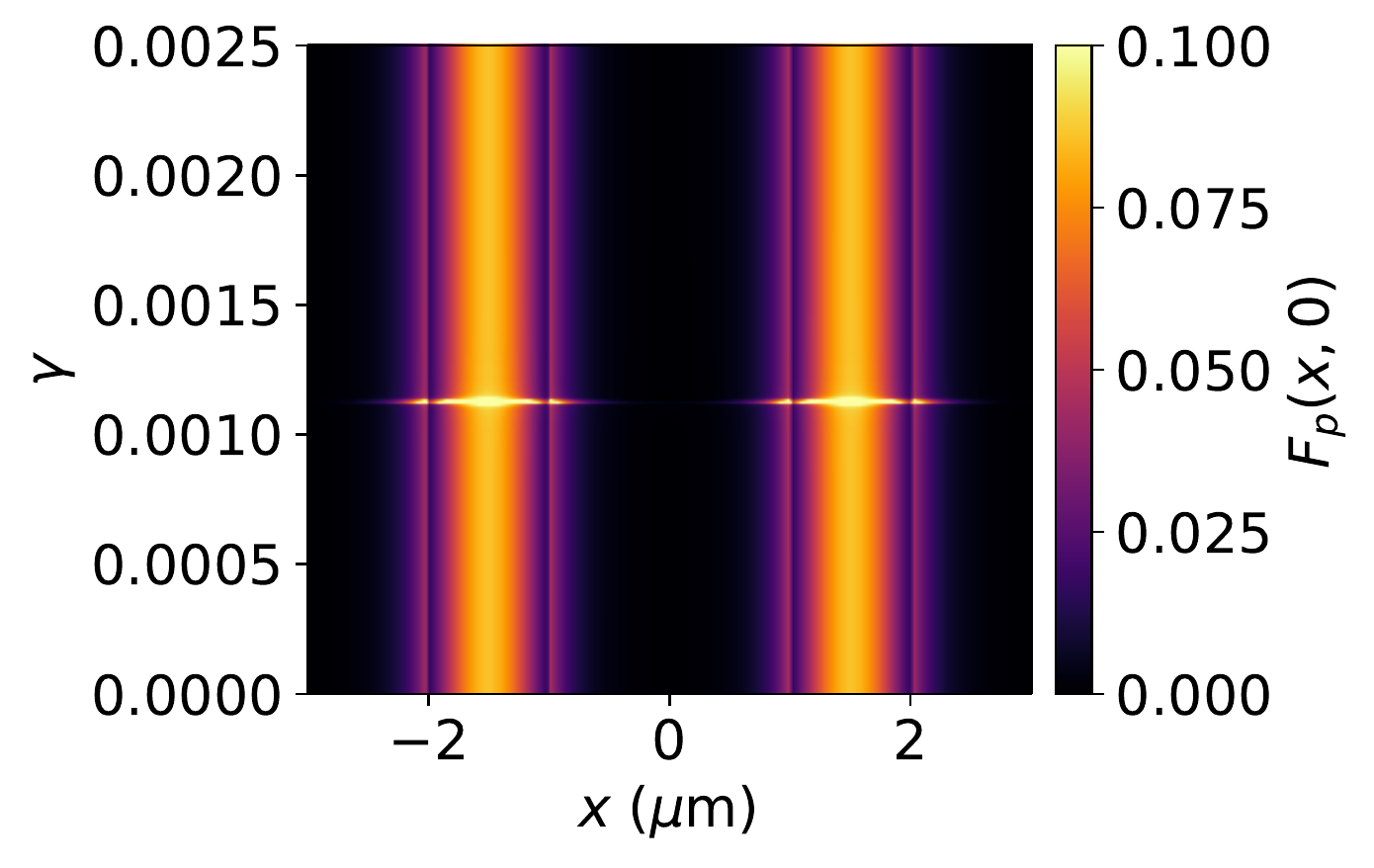}
  \caption{Distribution of the Purcell factor at the line
  $y=0$ as function of the emitter position $x$ and non-Hermiticity parameter~$\gamma$. Parameters of the coupled waveguide are given in the caption of Fig.~\ref{fig:Fp_in_plane}.}
  \label{fig:Fp_x_gamma}
\end{figure}

From Fig.~\ref{fig:Fp_in_plane} we conclude that the modal Purcell factor is symmetric in
(a) Hermitian regime as well as in (b) PT-symmetric and (c) PT-symmetry broken regimes.
In all three cases, the modal Purcell factor $F_p$ distribution is the same and finite (taking maximum value of approximately 0.085 in the middle of the waveguides)
despite the fact that both diagonal and off-diagonal terms experience enhancement as shown in Fig.~\ref{fig:Fp_gamma_y_is_0}.
According to the equations Eqs.~\eqref{eq:Fp_terms_sym} and \eqref{eq:Fp_terms_broken} this enhancement
is direct consequence of non-orthogonality.
Opposite signs and close absolute values of diagonal and off-diagonal terms observed in
Fig.~\ref{fig:Fp_gamma_y_is_0} result in cancellation of divergent terms in modal Purcell factor.
This explains small values of the modal Purcell factor and its independence on the non-Hermiticity parameter $\gamma$
demonstrated in Fig.~\ref{fig:Fp_x_gamma}.
Independence on the non-Hermiticity parameter
also confirms the analytical predictions given by Eqns.~\eqref{eq:Fp_sym} and \eqref{eq:Fp_broken}.
Note: a tiny spike observed near the EP is a numerical artefact.
It arises due to amplification of terms $F_{p,\mathrm{diag}}$ and $F_{p,\mathrm{off-diag}}$.

Such a behavior well agrees with the result obtained in Section~\ref{subsec:cmt}
utilizing the coupled-mode theory, namely, the numerically observed distribution of the modal Purcell factor is similar in Hermitian,
PT-symmetric, and PT-symmetry broken regimes.

\section{Summary and Outlook}
\label{sec:summary}

\begin{figure}[t!b!]
  \centering
  \includegraphics[width=0.7\linewidth]{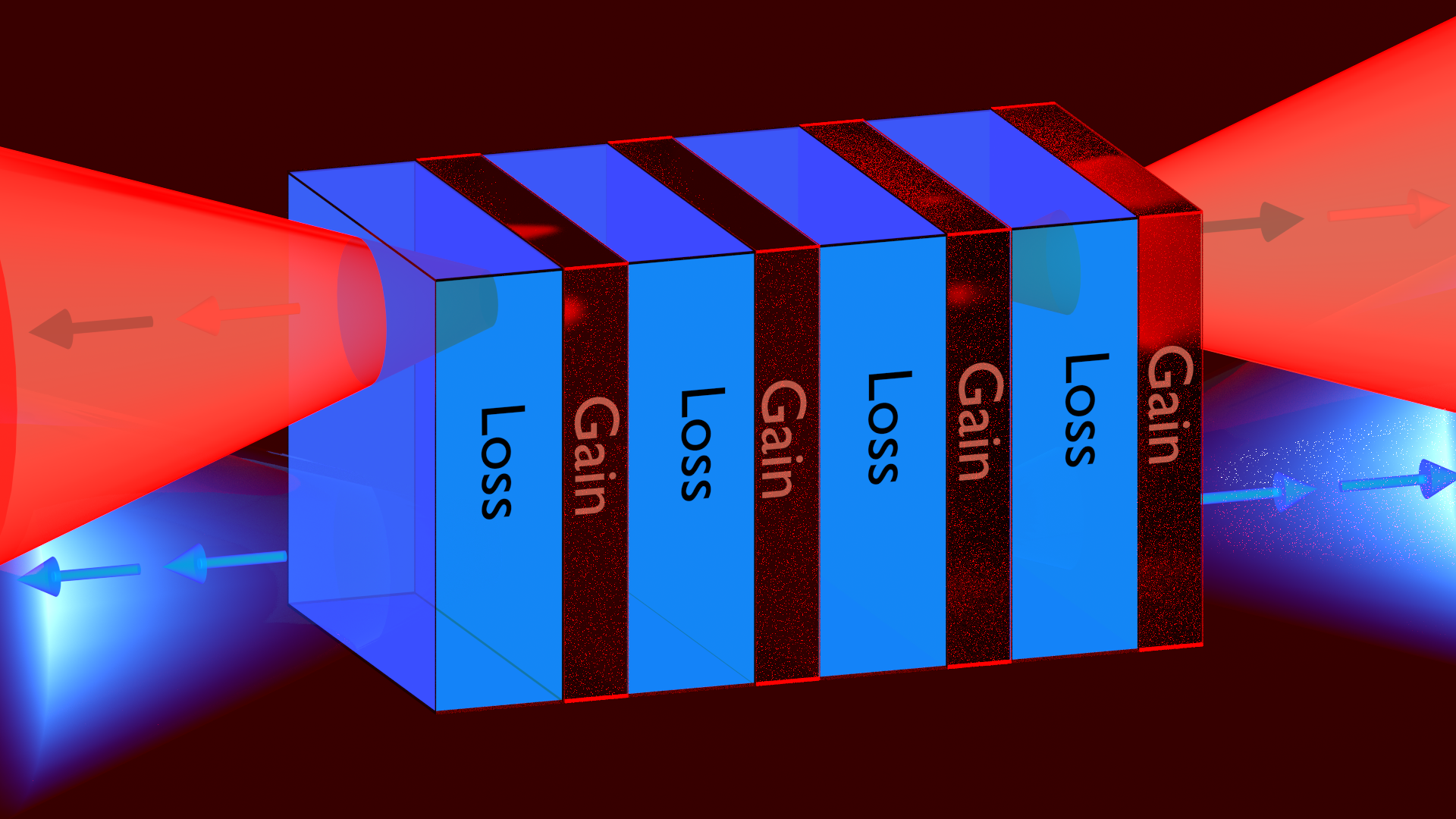}
  \caption{ Schematics of the proposed concept with N-periods multilayer waveguide core of alternating Loss and Gain media for on-chip lasing based on PT-symmetry effect. }
  \label{fig:Sect5}       
\end{figure}

To summarise, one of the challenges in integrated photonics, is to develop on-chip optical devices for efficient light manipulation finding its use in emerging applications such as data processing, quantum technologies, healthcare, security and sensing. Purcell effects in PT-symmetry can be utilized in variety of applications on a chip for instance for lasing. Lasing like behaviour  can be realised based on multilayer system releasing the pumped energy in the form of powerful pulses \cite{ref:novitsky2018pt}. Similar approach studied in \cite{ref:novitsky2018pt} can be implemented on a chip.
Figure \ref{fig:Sect5} shows the concept of Transmission and Reflection through the multilayered waveguide core, composed from Loss and Gain media.

Although efficient on-chip light manipulation can be achieved by engineering artificial materials (metamaterials) with unique optical permittivities and permeabilities, PT-symmetric photonics allows to tune the complex refractive index and control the interplay between the phase (real part of complex refractive index) and attenuation or loss (in case the imaginary part of complex refractive index is negative);
or amplification/gain in case the imaginary part of complex refractive index is positive.
The major advantage of PT-symmetric systems is to confine and guide light in coupled passive waveguides as was first shown in ref.~\cite{ref:guo2009}.
Then, the active fully PT-symmetric system with gain and loss was demonstrated using two coupled waveguides fabricated from Fe-doped \ce{LiNbO3} \cite{ref:ruter2010} in such a way that the transmission always appeared at the output of the active waveguide regardless of the input waveguide.
This effect is named non-reciprocal meaning that power oscillations between the coupled waveguides are asymmetric.
The degree of non-reciprocity in such nonlinear devices depends on the intensity of the signal.
However in \cite{ref:ruter2010}, Lorentz reciprocity still holds as long as no nonlinearity builds up.

PT-optomechanics is another interesting way to go and explore the interaction between the optical fields and mechanical option in PT-symmetric systems in presence of a quantum emitter.
In coupled mechanical resonators with optically induced loss and gain, a combination of nonlinear saturation and noise leads to preserved or weakly broken PT-symmetry, and a transition occurs from a thermal to a lasing state with small amplitude \cite{kepesidis2016symmetry,ozdemir2019parity}.

Systems with exceptional points, particularly, PT-symmetric systems are known to be able to enhance~\cite{ref:pick2017} and suppress~\cite{ref:akbarzadeh2019,ref:khanbekyan2020} the spontaneous emission rate in optical systems when operating near exceptional point.
Analysis of the spontaneous emission enhancement and coupling to the guided modes of the PT-symmetric coupled waveguide system shown that, interestingly, for this class of systems the modal enhancement factor (modal Purcell factor) does not depend on the non-Hermiticity even at the EP.

In conclusion, although the PT symmetry and non-Hermiticity in integrated photonics research has already established novel ways of utilizing gain, loss and their coupling to control light transport, there is still a room for new direction to go, when considering a Purcell effect in PT-symmetric waveguides.\\

\textbf{Acknowledgment}
AK acknowledges the support of Israel Science Foundation (ISF) Grant no. 2598/20.

\newpage
\bibliographystyle{spphys}
\bibliography{references}
\end{document}